\documentclass[pmlr]{jmlr}

\usepackage{longtable}

\usepackage{booktabs}
\usepackage[load-configurations=version-1]{siunitx} 

\usepackage{multirow}

\theorembodyfont{\upshape}
\theoremheaderfont{\scshape}
\theorempostheader{:}
\theoremsep{\newline}

\jmlrvolume{166}
\jmlryear{2022}
\jmlrworkshop{Holistic Evaluation of Audio Representations}

\title[Learning Audio Representations with MLPs]{Learning Audio Representations with MLPs}

\author{
  \Name{Mashrur M. Morshed}$^{1}$  \Email{mashrur@intelligentmachin.es}\thanks{alt. email: mashrurmahmud@iut-dhaka.edu}\\
  \Name{Ahmad Omar Ahsan}$^{1}$ \Email{omar@intelligentmachin.es}\thanks{alt. email: ahmadomar@iut-dhaka.edu}\\
  \Name{Hasan Mahmud}$^{2}$ \Email{hasan@iut-dhaka.edu}\\
  \Name{Md. Kamrul Hasan}$^{2}$ \Email{hasank@iut-dhaka.edu}\\
  \AND
  \addr $^{1}$ Intelligent Machines Limited \\
  \addr $^{2}$ Systems \& Software Lab, Dept. of CSE, Islamic University of Technology
 }

\editor{Editor's name}

\begin{document}

\maketitle

\begin{abstract}
In this paper, we propose an efficient MLP-based approach for learning audio representations, namely timestamp and scene-level audio embeddings. We use an encoder consisting of sequentially stacked gated MLP blocks, which accept 2D MFCCs as inputs. In addition, we also provide a simple temporal interpolation-based algorithm for computing scene-level embeddings from timestamp embeddings. The audio representations generated by our method are evaluated across a diverse set of benchmarks at the Holistic Evaluation of Audio Representations (HEAR) challenge, hosted at the NeurIPS 2021 competition track. We achieved first place on the Speech Commands (full), Speech Commands (5 hours), and the Mridingham Tonic benchmarks. Furthermore, our approach is also the most resource-efficient among all the submitted methods, in terms of both the number of model parameters and the time required to compute embeddings.

\end{abstract}
\begin{keywords}
Audio Representations, Embeddings, MLP, Interpolation, Representation Learning
\end{keywords}

\section{Introduction}
\label{sec: introduction}

Multilayer Perceptrons (MLP) are densely connected (also called fully connected) feed-forward neural networks, where all the neurons in an arbitrary hidden layer are connected to every other neuron in the subsequent hidden layer \citep{multilayerperceptron}. MLPs are widely considered as \textit{classical} deep neural networks.

Over the past decade, researchers have opted for other neural network variants, such as Convolutional Neural Networks (CNN) or Recurrent Neural Networks (RNN), because of some shortcomings of classical MLPs. For instance, CNNs have the very useful translation-invariance property, which make them uniquely suitable for Computer Vision and spatial pattern recognition tasks. Furthermore, because convolutional kernels are windowed across inputs, they implement a form of weight sharing \citep{cnninvariance}. This makes them many times more efficient in terms of computation and parameters over classical MLPs, in addition to their better recognition performance. RNNs are similarly preferred for tasks with sequential or time-based features, or tasks where it is necessary to model temporal dependency.

However, MLPs have lately seen a resurgence in the field of deep learning, particularly in vision. Recent research \citep{mlpmixer,gmlp} propose efficient ways of using MLPs to learn spatial and channel information in a way that makes them competitive with CNNs and other state-of-the-art approaches (such as vision transformers \citep{vit}).

In this paper, we show that MLPs can be used to learn useful audio representations from Mel Filter Cepstrum Coefficients (MFCC), which are 2D signals like images. While images are a 2D function of two spatial dimensions, MFCCs are instead a function over time and mel-frequencies. The channel and spatial operations of MLPs on images can be translated to frequency and temporal operations on MFCCs, allowing us to learn high-level audio representations which can be used on downstream tasks. We provide two types of learned representations: timestamp-level embeddings and scene-level embeddings.

The scene-level embeddings are generated from the timestamp-level embeddings, using an iterative linear interpolation-based algorithm over the temporal dimension. We perform a study of three scene-embedding algorithm approaches, and show that in general, interpolation-based algorithms do perform better than naive averaging approaches, and that the choice of interpolation algorithms have a subtle but clear effect on downstream task performance.

Our method was evaluated at the Holistic Evaluation of Audio Representations (HEAR) challenge \citep{turian2022hear}, hosted at the NeurIPS 2021 competition track. The challenge evaluated submissions across a highly diverse set of nineteen tasks (derived from sixteen datasets). As a part of the challenge, we provided a Python package following a common API developed by the organizers. This means that our work is reproducible, easy to use in future research and applications, and also extensively evaluated on diverse downstream audio problems, which clearly highlights both the strengths and limitations of our approach.

We can summarize our contributions as following:

\begin{enumerate}
    \item We propose an all-MLP approach for learning timestamp and scene-level embeddings for audio, which shows promising results after extensive evaluation on a diverse set of audio benchmarks. Our method is the most efficient among all submissions in terms of both model parameters and embedding generation time.
    
    \item We propose an iterative linear interpolation-based algorithm for generating scene-level embeddings from timestamp-level embeddings.
    
    \item We perform several ablation experiments, with which we can improve some of our existing results, and even achieve results better than the first place score on certain tasks.
    
\end{enumerate}


The rest of this paper is organized as follows: in Section \ref{sec: background} we provide a brief background on some of the related concepts, as well as a review of relevant related works. In Section \ref{sec: method} we describe our proposed method and approach to the HEAR challenge. We observe our results on HEAR in Section \ref{sec: results} and analyze those results from various perspectives in Section \ref{sec: analysis}. We then describe three types of ablation experiments and their respective results in Section \ref{sec: ablation}, followed by our conclusion in Section \ref{sec: conclusion}.


\section{Background}
\label{sec: background}

\subsection{Low-Level Audio Representations}
By low-level, we refer to representations which have not been learned through any supervised or unsupervised training method. The lowest-level audio representation is naturally the digital audio signal, directly obtained from the raw analog audio or sound. Generally, digital audio can be represented as a numeric array of length $t \times sr$, where $t$ is the time duration of the audio and $sr$ is the sampling rate. This is a very high dimensional representation, and thus requires additional processing in order to be used in discriminative tasks. More useful low-level representations with lower dimensionality can be obtained by applying handcrafted transforms, such as log-Mel Spectrograms \citep{logmelspec}, Mel Frequency Cepstrum Coefficients (MFCC) \citep{mfcc}, Constant Q-Transform (CQT) \citep{cqt}, Variable Q-Transform (VQT) \citep{vqt}, and Gammatone Filterbanks \citep{gammatone}. High-level audio representations (embeddings) are learned by training models which take the low-level representations as inputs, and produces vectors or feature-maps. In our work, our model learns audio embeddings by using MFCCs as inputs.

\subsection{Learned Audio Representations}
Throughout the past decade, numerous methods have been explored to learn audio embeddings (which can be considered as high-level audio representations). 

PANNs (Pretrained Audio Neural Networks) \citep{panns} proposed a variety of audio CNN models with Mel Spectrogram inputs, pretrained at a large scale on about 5000 hours of audio from the AudioSet dataset \citep{audioset}. In the original PANNs paper, the authors evaluated its performance across six audio pattern recognition tasks, including audio tagging and sound event detection. \citet{ast} proposed the AST (Audio Spectrogram Transformer), which uses DEIT (Data-efficient Image Transformers) \citep{deit} pretrained on ImageNet \citep{imagenet} and further pretrains on AudioSet, showing good performance on several tasks. In a later work, \citep{ssast} propose a self-supervised way to train AST, showing better downstream task performance.  SoundNet \citep{soundnet} learns audio representations from unlabelled videos by exploiting the synchronization between audio and video.

\citet{wav2vec}, in their work Wav2Vec, explored an unsupervised pre-training approach for speech recognition by learning representations from large amounts of raw, unlabelled audio. In their later work Wav2Vec2.0 \citep{wav2vec2.0}, the authors explore a self-supervised method where they use a single hour of labelled data and large amounts of unlabelled data to outperform existing speech recognition methods.

\citet{tripletloss} used a triplet loss-based embedding and a kNN classifier to significantly improve the performance of convolutional networks on Speech datasets. \citet{multicontrast} use contrastive learning to jointly learn audio representations from raw audio and spectrograms. CREPE \citep{crepe} proposes a CNN-based approach for learning pitch-sensitive representations.

\citet{decoar} propose a representation learning approach where reconstruction is done from past and future context frames; the learned deep contextualized representations reduce the need for labelled data. TERA \citep{tera} (Transformer Encoder Representations from Alteration) is a self-supervised speech pretraining method that trains transformer encoders on a large amount of unlabelled speech data. TERA is evaluated on several downstream tasks, such as phoneme classification, keyword spotting, speaker recognition, and speech recognition. Lastly, HuBERT (Hidden-unit BERT) \citep{hubert} uses a 1 billion parameter model and an offline clustering step to match or improve on the performance of Wav2Vec2.0 on several speech tasks.

\subsection{Timestamp and Scene-level Embeddings}
\label{sec: timestamp-scene-embeddings}

\paragraph{Timestamp-level Embeddings} These embeddings either represent a particular time \textit{instance} in the audio, or a very short period centered around that particular time instance. Each timestamp embedding is in practice a numeric vector of certain size, which we refer to as the timestamp embedding size ($E_{T}$).

From a single audio clip, we can generate a variable number of timestamp embeddings ($N_{T}$), depending on the duration of the input audio and the hop length or stride of the model. $N_T$ is roughly equivalent to the length of the audio divided by the hop length.

Timestamp-level embeddings for a single audio are thus of the shape $N_T \times E_T$. As these embeddings represent specific, temporally ordered instances in the audio, they can be used for tasks like Sound Event Detection.

\paragraph{Scene-level Embeddings} On the other hand, scene-level embeddings represent the entire input audio. These embeddings are numeric vectors which remain fixed in size even with input audios of variable length. Naturally, there can only be a single scene-level embedding for a single audio. The length of the scene embedding vector is what we refer to as the scene embedding size ($E_S$). Since scene-level embeddings represent the entire audio, they are mainly useful in various audio classification tasks.

\subsection{Transformers}

Transformers \citep{transformer} have been a highly disruptive innovation in deep learning. They have shown remarkable performance in numerous Natural Language Processing problems \citep{bert,roberta} and can currently be considered as the de facto approach to NLP \citep{huggingface}, replacing recurrent neural networks.

With the advent of Vision Transformers (ViT) \citep{vit}, they have also excelled in a large variety of Computer Vision problems \citep{transformersinvision}. A high-level reason for this is that transformers solve the problem of long-term dependencies, a problem not limited to the NLP domain; modelling long-term dependencies is also crucial to improving performance in many vision problems.

Transformers are also gaining predominance in the audio field. There are methods such as TERA \citep{tera}, Conformer \citep{conformer} (convolution-augmented transformers, used in speech recognition), and then ViT-like approaches such as the Keyword Transformer (KWT) \citep{kwt} and the Audio Spectrogram Transformer (AST) \citep{ast}. In recent self-supervised audio representation learning methods, transformer-based encoders have seen much use alongside convolutional or convolutional-recurrent encoders \citep{audiossl}. 

\subsection{Gated MLPs (gMLP)}
\label{related-work:gmlp}
A recent group of works, the MLP-Mixer \citep{mlpmixer}, gated MLPs (gMLP) \citep{gmlp}, ResMLP \citep{resmlp}, as well as the work of \citet{melas}, provide evidence that self-attention, a core component in transformers and the ViT, may not be solely responsible for their remarkable performance on vision tasks. These models are attention-free and are solely comprised of MLPs, showing that the patch-embedding process they share with ViTs may be the crucial component. We mainly focus on gMLP here, as our work is closely related with theirs.

Like ViT, the gMLP is a patch-based model. It accepts patches of images as inputs, which are first projected to a higher dimension through a patch embedding layer. Then these patch embeddings are passed through a stack of gMLP blocks, where each block is effectively a pair of channel-wise linear projections, separated by a spatial linear projection. 

Keyword-MLP (KW-MLP) \citep{kwmlp} is a much smaller gMLP-like model used for Keyword Spotting on Speech Commands \citep{speechcommands}. Instead of patches of an image, KW-MLP accepts a 2D MFCC without patching. The frequency dimension of MFCC is projected to higher dimensions, followed by a variant of the gMLP block which applies a pair of linear projections across the frequency dimension, separated by a linear projection across the temporal dimension. While the original gMLP model used on ImageNet has three variants with 6M, 20M and 73M parameters, KW-MLP has only 0.4M parameters.

We use the KW-MLP as an encoder in our approach, due to its very efficient size and some of its unique properties, which we will discuss in Section \ref{kwmlp-encoder}.

\subsection{Holistic Evaluation of Audio Representations (HEAR)}

From some of the existing works discussed above, it can be observed that audio representation learning methods are generally evaluated on a limited set of downstream tasks, usually containing speech recognition. This is a problem, because the primary use of learned audio representations is to obtain better performance in solving downstream tasks, particularly low-resource tasks for which there aren't much available data. Because of the lack of extensive evaluation and the generally homogeneous type of downstream tasks existing models are evaluated on, we do not have a good idea of what representation would excel on a new, unknown task.

The HEAR challenge \citep{turian2022hear} addressed this very concern, by evaluating methods on nineteen different tasks, derived from sixteen different datasets. Submitted audio representations (timestamp and scene-level embeddings) were used to train a very shallow fully connected network, without finetuning the embedding model at all. This allowed empirical evaluation of the generalization capability of various learned audio representations.

\begin{figure}[ht]
\centering
\includegraphics[width=0.6\linewidth]{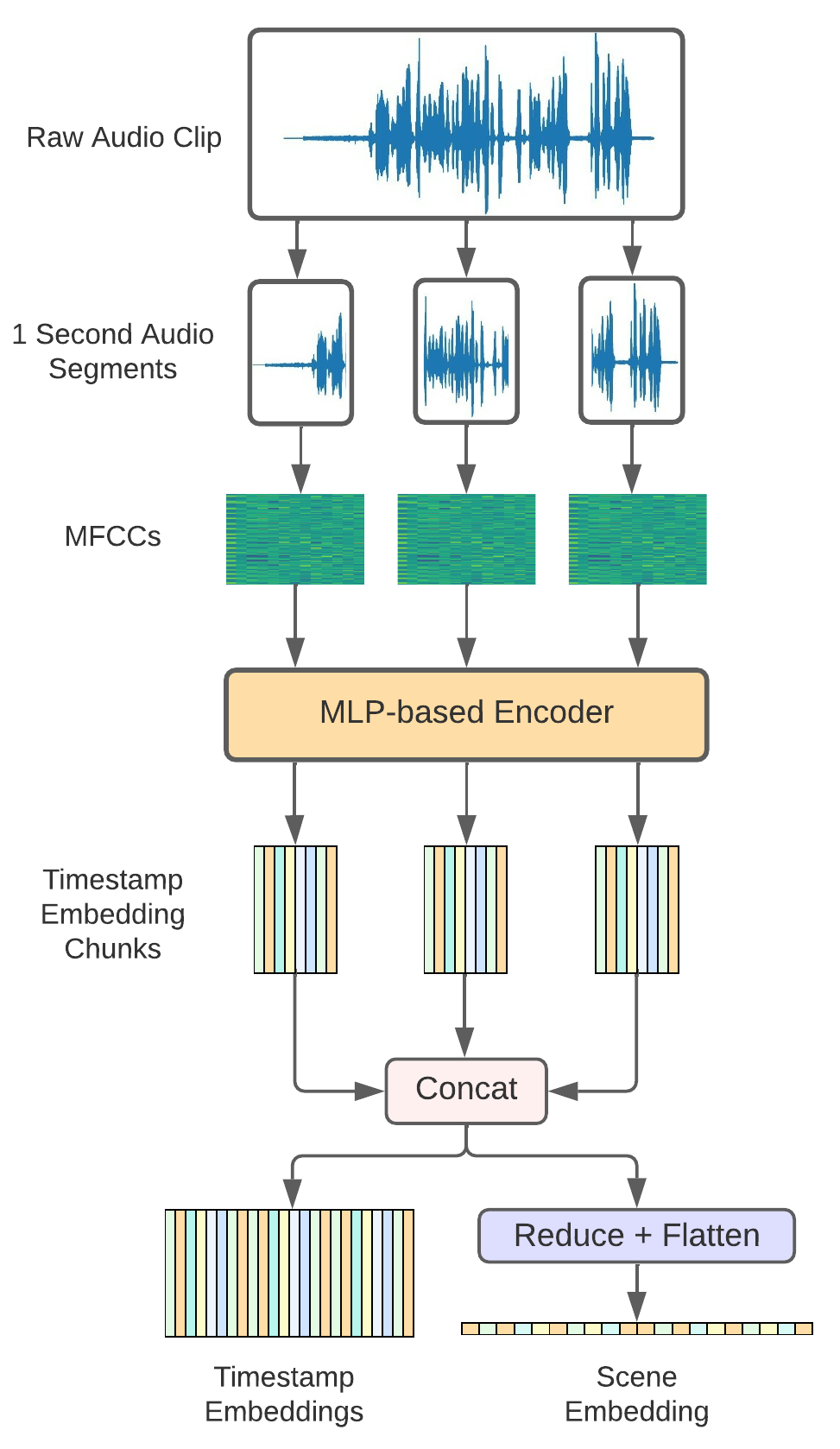}
\caption{Overview of our approach. Each audio clip is divided into 1 s audio segments, which are then converted to MFCCs and passed to the MLP-based encoder.}
\label{fig:overview}
\end{figure}

\section{Method}
\label{sec: method}

We approached the HEAR challenge by first training the KW-MLP model on the Google Speech Commands V2-35 dataset \citep{speechcommands}. It is to be noted that one of the open datasets in the HEAR challenge also featured Speech Commands, but a different variant: the Speech Commands V2-12 dataset. We trained our encoding model on the 35-class variant of the dataset with the aim of learning more discriminative features.

After training, we detached the linear classification head from the tail end of the model, and froze its weights. This frozen model, comprised solely of MLP layers, is our encoder. A wrapper was added around the encoder to handle audio processing, and subsequent generation of scene-level embeddings from timestamp embeddings.

Lastly, we implemented the common API \citep{commonapi} proposed by the HEAR organizers, and ensured that our model wrapper was compatible with the API standards. Afterwards, the HEAR organizers used the HEAR Evaluation Kit \citep{heareval} to evaluate our submission by first generating embeddings and then training shallow neural networks, on the nineteen tasks \citep{turian2022hear}. For the results we mention in our ablation studies (Section \ref{sec: ablation}), we perform the evaluation experiments ourselves using the same open source evaluation kit.

\subsection{Audio Preprocessing}

As per the Common API established by the HEAR organizers , each model is required to perform at one of 16000, 22050, 32000, 44100 or 48000 Hz sampling rate. Our model accepts audio of 16000 Hz sampling rate, so any input audio needs to first be resampled to that rate.

Next, the Keyword-MLP model which we use as our encoder or feature extractor operates on MFCCs of 1 s audio. So after resampling, any arbitrary input audio needs to be divided into 1 s long segments. For fractional durations, we simply perform ceiling padding at the end of the audio to lengthen it to the next s (e.g. for 4.5 s input audio, we pad it to 5 s).

If an input audio has the duration $b$ s (after appropriate padding) and a sampling rate $sr$, then we obtain a tensor of shape $(b, sr)$, where $b$ is the batch size (i.e. we have a batch of 1 s audios). We then compute the MFCC for this batch of 1 s audio tensors in parallel with \texttt{nnAudio} \citep{nnaudio}, an audio processing toolbox that uses PyTorch-based 1D CNNs as its backend. This allows us to leverage GPU-parallelism when extracting features on CUDA runtimes.

We compute MFCCs with a hop size of 10 ms (0.01 s), a window length of 30 ms (0.03 s), and 40 MFCC bins. The encoder model subsequently accepts inputs of shape (40, 98). All audio preprocessing settings are detailed in Section \ref{appendix: kwmlp-hparams}.

\subsection{KW-MLP Encoder}
\label{kwmlp-encoder}

Keyword-MLP (KW-MLP) \citep{kwmlp} takes MFCCs of 1 s audio clips as input. Let us assume $X \in {\mathbb{R}}^{F \times T}$ is an input MFCC, where $F$ and $T$ are the mel frequency bins and time-steps respectively. As mentioned earlier in Section \ref{related-work:gmlp}, KW-MLP does not require patching its 2D input, unlike ViT and gMLP. Technically, $X$ is to be divided into $T$ patches of shape $F\times{1}$, getting $X_{0} \in {\mathbb{R}}^{T \times F} $(thus in practice, patching is actually not required, and is equivalent to a transpose).

Using a linear projection matrix $P_{0} \in {\mathbb{R}}^{F \times d}$, $X_{0}$ is projected to a higher dimension $d$, which is the embedding size of the model (and also the timestamp embedding size). We obtain $X_{E} \in {\mathbb{R}}^{T \times d}$ from this operation.

\begin{equation}
    X_{E} = X_{0}P_{0}
\end{equation}

$X_{E}$ can be considered as a frequency domain patch-embedding, as it is a result of directly projecting the frequency dimension of the input MFCC.

\begin{figure}[ht]
\centering
\includegraphics[width=0.6\linewidth]{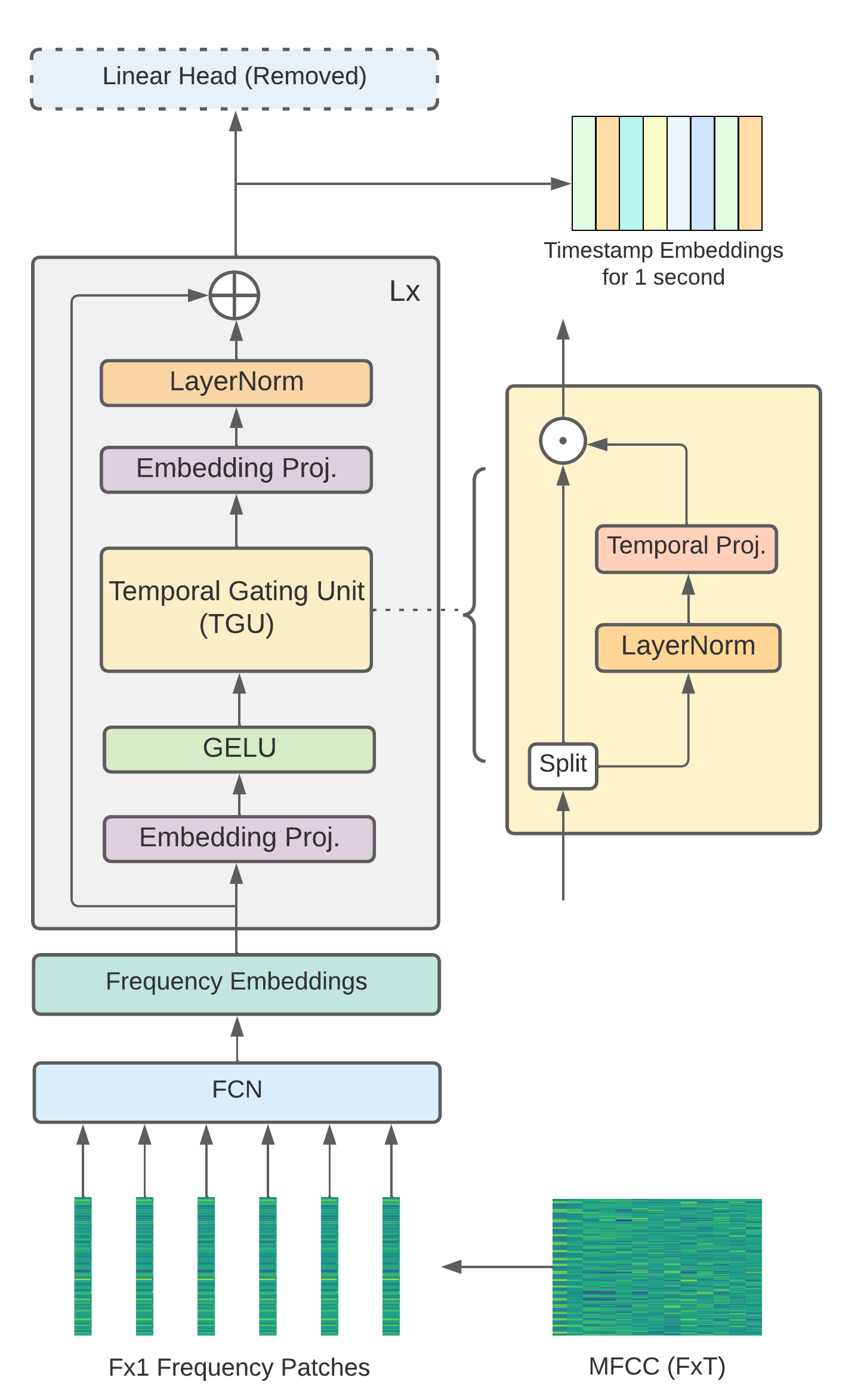}
\caption{Keyword-MLP architecture. It contains $L=12$ consecutive gated-MLP blocks, and operates on patches from MFCCs of 1 s audio segments.}
\label{fig:kwmlp}
\end{figure}

The obtained $X_{E}$ is passed through $L$ consecutive, identical gated-MLP (gMLP) blocks. These blocks can be seen as a pair of projections across the frequency embedding dimension, separated by a projection across the temporal dimension. The block can also be formulated with the following set of equations:

\begin{equation}
\begin{split}
 Z  & =  \sigma (X_{in} U) \\
\tilde Z  =  g(Z) & = g([Z_{r} Z_{g}]) = Z_{r} \odot (Z_{g}^{T}G)^{T}  \\
X_{out}  & = \tilde ZV \oplus X_{in}
\end{split}
\label{eqn:gmlp}
\end{equation}

Here, $\sigma$ represents the GELU activation function. $U \in {\mathbb{R}}^{d\times{D}}$ and $V \in {\mathbb{R}}^{D/2\times{d}}$ are the linear projection matrices which project from $d$ to the higher dimension $D$, and $D/2$ back to $d$ respectively. $Z \in {\mathbb{R}}^{T\times{D}}$ is split to $Z_r$ (residual) and $Z_g$ (gate) $\in {\mathbb{R}}^{T\times{D/2}}$ respectively,  while $G \in {\mathbb{R}}^{T\times{T}}$ performs linear projection on $Z_g$ across the temporal dimension. Each projection is effectively done with a matrix multiplication, and in practice, can be implemented as a \texttt{Dense} layer. The operation $(Z_{g}^{T}G)^{T}$, the temporal projection, requires two additional transpose operations. This can be simplified as a pass through a \texttt{Conv1D(T, T, 1)} layer.

(Note that $A^{T}$ signifies a transposed matrix, and $T$ in this context is not the time dimension).

The overall architecture of the encoder can be observed in Figure \ref{fig:kwmlp}. The embedding size of KW-MLP $d$ (also the timestamp embedding size, $E_T$), is 64, while the embedding projection size $D$ is 256. We used $L = 12$ blocks like the one shown in equation \ref{eqn:gmlp} and in Figure \ref{fig:kwmlp}, in our submission to the HEAR challenge. For later ablation experiments (Section \ref{ablation: shallower-kwmlp}) we used fewer blocks (such as 10 and 8). All other hyperparameters and settings are shown in Section \ref{appendix: kwmlp-hparams}. It is to be noted that the pretraining task of KW-MLP (training on Speech Commands V2-35) can be done very quickly, as fast as about two hours on Google Colab.

KW-MLP$_L$ denotes a KW-MLP model with $L$ blocks. The default model that we simply refer to as KW-MLP is technically KW-MLP$_{12}$.

\subsection{Interpolation-based Scene Embeddings}

\begin{figure}[ht]
\centering
\includegraphics[width=0.8\linewidth]{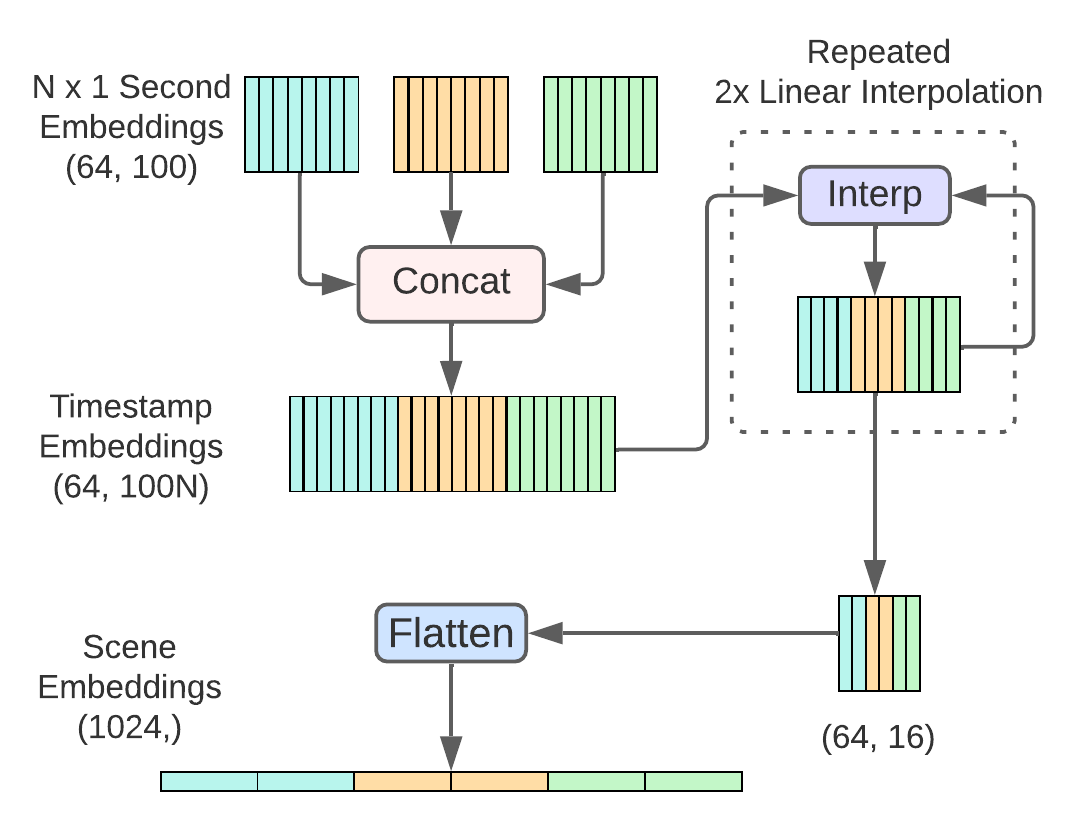}
\caption{After concatenation, timestamp embeddings are repeatedly downsampled by linear interpolation till target size is achieved. Then scene embeddings are obtained by flattening.}
\label{fig:embedding}
\end{figure}

We use a hop size of 10 ms. Thus, for each 1 s audio segment, our model outputs embeddings of size $(100, 64)$, where 64 is the timestamp embedding size, $E_T$ (Section \ref{sec: timestamp-scene-embeddings}). We concatenate these embeddings along the time axis to obtain the output timestamp embeddings: for an $N$ s input audio, we would have embeddings of shape $(N_{T}, E_{T}) = (100 * N, 64)$. 

This variable-sized timestamp embedding further needs to be reduced to a fixed-size scene-level representation. We do this by downsampling the timestamp embedding across the time axis. The scene embedding size $E_{S}$ that we use is 1024, which is equivalent to $16 \times 64$. So, we need to reduce $N_{T}$ to $N_{S}$ (where $N_S = 16$). 

We use linear interpolation across the time axis to do this. However, instead of directly downsampling from $N_{T}$ to $N_S$, we do interpolation multiple times, each time reducing
$N_{T}$ to $N_{T}/2$. The number of required interpolations can be calculated by the following expression:

\begin{equation}
    N_{interp} = \left \lceil \log_{2} \left ( \frac{N_{T}}{N_{S}} \right ) \right \rceil
    \label{eqn:num_interp}
\end{equation}

\begin{figure}[ht]
\centering
\includegraphics[width=\linewidth]{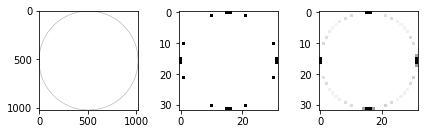}
\label{fig:circles}
\caption{(left) 1 unit thick circle of diameter 1024 (middle) Directly downsampled to 32x (right) Iteratively 2x downsampled to 32x}
\end{figure}

The primary rationale behind iteratively interpolating $N_{interp}$ times is to better conserve the embedding features when downsampling by a large factor. PyTorch's \texttt{F.interpolate} shows unsafe behaviour when directly downsampling by a large factor, as can be seen from Figure \ref{fig:circles}. Alongside PyTorch, some common Python image and array processing libraries exhibit similar behaviour \citep{twitter}.

In Section \ref{ablation: interp}, we perform an ablation study comparing this iterative interpolation with regular interpolation and averaging.

\section{Results}
\label{sec: results}

For the sake of reducing redundancy, we do not show full comparative results for each of the nineteen tasks. We instead only show the result of our approach and a comparative ranking in Table  \ref{tbl:result}.

The full results comparing every team submission per task can be viewed at the competition website \citep{hearresults} and in \citet{turian2022hear}. To re-clarify, our submitted Python package and overall submission is called \texttt{kwmlp}, and our team name is \texttt{ID56-SSL}. KW-MLP is the name of the encoder model, described in Section \ref{kwmlp-encoder}.
 
\begin{table*}[ht]
\centering
\caption{Results of \texttt{kwmlp} on the nineteen tasks at HEAR 2021 \citep{turian2022hear}}
\label{tbl:result}
\begin{tabular}{ |c|c|c|c|  }
\hline
    Task & Metric & Result & Ranking \\
\hline
    Speech Commands (V2-12) Full               & Accuracy   & \textbf{0.978} & \textbf{1st}  \\
\hline
    Speech Commands (V2-12) 5H                 & Accuracy   & \textbf{0.976} & \textbf{1st}  \\
\hline
    \multirow{2}{*}{NSynth Pitch 5H}   & Pitch Acc  & 0.440 & 11th \\

                                       & Chroma Acc & 0.480 & 11th\\
\hline
    \multirow{2}{*}{NSynth Pitch 50H}   & Pitch Acc  & 0.605 & 14th \\

                                       & Chroma Acc & 0.648 &  14th \\
 \hline
    \multirow{2}{*}{DCASE 2016 Task 2}   & Event Onset FMS  & 0.518 & 18th \\

                                       & Segment Error Rate & 0.527 &   21st \\
\hline
    \multirow{2}{*}{Beehive States}   & AUCROC  & \textbf{0.760} & \textbf{3rd} \\
                                       & Accuracy & \textbf{0.552} & \textbf{5th}\\
\hline
    Beijing Opera Percussion  & Accuracy  & 0.911 & 24th \\
\hline
    CREMA-D & Accuracy & 0.424  & 25th \\
\hline
    ESC-50  & Accuracy  & 0.367 & 25th \\
\hline
    \multirow{2}{*}{FSD50K}   & mAP  & 0.187 & 22nd \\
                                       & d' & 1.511 & 21st \\    
 \hline
    Gunshot Triangulation  & Accuracy  & \textbf{0.932} & \textbf{7th} \\
\hline
   GTZAN genre & Accuracy & 0.554 & 27th \\
\hline
    GTZAN Music Speech & Accuracy & 0.889 & 27th \\
\hline
    LibriCount  & Accuracy  & 0.451 & 27th \\
\hline
    \multirow{2}{*}{Maestro 5H}   & Note Onset FMS  & 0.648 & 11th \\
                     & Note Onset w/ Offset FMS & 0.0210 & 11th \\ 
\hline
    Mridingham Stroke  & Accuracy  & \textbf{0.969} & \textbf{5th} \\
    
\hline
    Mridingham Tonic  & Accuracy  & \textbf{0.942} & \textbf{1st} \\
\hline
    \multirow{2}{*}{Vocal Imitations}   & mAP  & 0.056 & 26th \\
                     & Accuracy & 0.049 & 26th \\ 
 \hline
    Vox Lingua Top 10 & Accuracy  & 0.1812 & 24th \\
\hline
\end{tabular}
\end{table*}

\section{Analysis}
\label{sec: analysis}

In this section, we analyze the observed results from Table \ref{tbl:result} from multiple perspectives.

\subsection{Tasks With Good Performance}
\texttt{kwmlp} ranks first on both the complete Speech Commands as well as the shorter and more challenging 5 hour variant. We do not consider this an unexpected result, as the KW-MLP model that we use as an encoder was trained for keyword spotting on Speech Commands V2-35. It is to be noted however that the tasks aren't quite the same; the HEAR challenge evaluated the V2-12 variant of the dataset.

\texttt{kwmlp} also performs remarkably well on the Mridingham Stroke and Mridingham Tonic tasks \citep{mridingham}. At the very least, we believe it is a very interesting outcome that the embeddings learned by an extremely small model, on a rather small dataset, can outperform much larger models pretrained on larger audio corpuses on a secret task. This provides evidence that audio representations learned by isotropic, all-MLP models require further research.

In our ablation experiments (Section \ref{ablation: shallower-embedding-kwmlp}) we obtain improved results for various tasks, including better than first place results on Gunshot Triangulation \citep{gunshot} and Mridingham Stroke and Tonic \citep{mridingham}.

\subsection{Tasks With Poor Performance}
Because of supervised pretraining on Speech Commands, we believe our model has learned to be \textit{temporally shift invariant}. As a result, performance on FSD50K \citep{FSD50k} and DCASE \citep{dcase} suffers. In Section \ref{temporal-shift}, we analyze this by visualizing the model weights. Our method is also pitch and tone agnostic, thus suffering on tasks like GTZAN \citep{gtzan}, ESC-50 \citep{esc50}, etc.

On language-based dataset such as Vox Lingua \citep{voxlingua}, our method did not perform well since our model was trained only English audio. Again, training on Speech Commands also means not discriminating between voices, which results in reduced performance on Vocal Imitations \citep{vocalimitations}. Performance in these two tasks are generally sub-par among the submissions, as most models are pretrained on English speech and may be invariant to vocal identity.

\subsection{Audio Duration} The task containing the longest audio clips in the HEAR challenge was Beehive States \citep{beehive}, at ten minutes per clip. On the other side, there are tasks with much shorter audio as well, such as Speech Commands, containing 1 s audio clips. We can see from Table \ref{tbl:result} that duration does not have much impact on results.

\subsection{Performance on Smallest Dataset}
Learned representations are particularly important for approaching novel tasks with scarce data. The task containing the fewest number of data instances is Gunshot Triangulation \citep{gunshot}, containing only 88 audio clips. The original \texttt{kwmlp} performs reasonably well on this task, while in one of our ablation experiments (Section \ref{ablation: shallower-embedding-kwmlp}, Table \ref{tbl:kwmlp-representation}), we obtain results better than the first place result in the competition.

\subsection{Comparative Model Efficiency}

Apart from accuracy, it is also important to consider the size and comparative resource-efficiency of various models in the HEAR benchmark. For instance, for the Beehive States task \citep{beehive}, only 11 submissions out of 28 could be evaluated, as the rest exceeded the GPU memory and time restrictions established by the organizers. Beehive States is the task with the longest audio clips among all HEAR tasks, at ten minutes long; thus we assume some of the larger models scaled exponentially and exceeded resource limits.

\texttt{kwmlp} has a total of only 0.424 million parameters. For reference, the immediate model after \texttt{kwmlp} in terms of size is \texttt{YAMNet}, at 3.8 million parameters ($\sim 9\times$), while the largest models, combinations of HuBERT, Wav2Vec2.0, and CREPE in a variety of ways \citep{ntu}, reached 1339 million parameters (over 3000$\times$ larger). 

\subsection{Temporal Shift Invariance}
\label{temporal-shift}

\begin{figure}[ht]
\centering
\includegraphics[width=\linewidth]{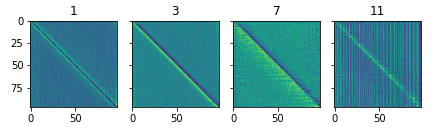}
\caption{Visualization of the linear temporal projection matrix ($G$ from equation \ref{eqn:gmlp}) at blocks of different depths (1, 3, 7, 11).}
\label{fig:temporal-shift-inv}
\end{figure}

We can observe an interesting phenonmenon if we visualize the weights of the temporal linear projection (in Figure \ref{fig:temporal-shift-inv}): that they resemble toeplitz matrices.

This implies that the encoder model has learned a form of shift invariance in its supervised pretraining task. To be precise, this is temporal shift invariance, as these weight matrices are responsible for linear temporal projection. This makes sense upon reviewing the nature of Speech Commands classification, which is the task KW-MLP is trained on. In an arbitrary input audio, the target speech keyword can appear at various positions, such as at the start, the middle, or the end. The words can also have different time durations, depending on the utterance style. Naturally, a model aiming to solve the Speech Commands task must learn temporal shift invariance. As a consequence, it becomes expected for our method to not perform well on timestamp embedding tasks, which we can observe from the results on DCASE.


\section{Ablation}
\label{sec: ablation}
For our ablation studies, we performed experiments on eight tasks with relatively small datasets, which enabled quick experimentation and did not require much compute resources.

\subsection{Understanding the Impact of Scene Embedding Algorithms}
\label{ablation: interp}

For HEAR 2021, in \texttt{kwmlp}, we used iterative linear interpolation to generate scene-level embeddings from timestamp-level embeddings.  We were curious to observe the effect of iterative interpolation as opposed to regular (single) interpolation. We also made a comparison with a mean-based algorithm, which makes $N_S$ (16) groups (in order) along the time dimension and takes the average per group before flattening. The results of the three algorithms is compared in Table \ref{tbl:embeddings}.

\begin{table*}[ht]
\centering
\caption{Impact of three different scene embedding algorithms}
\label{tbl:embeddings}
\begin{tabular}{ |c|c|c|c|c|  }
\hline
    \multirow{2}{*}{Task} & \multirow{2}{*}{Metric} & Group & \multicolumn{2}{c|}{Interpolation} \\
    \cline{4-5}
    & &  Mean & Single & Iterative \\
\hline
    Beijing Opera Percussion & Accuracy & 0.894 & 0.902 & \textbf{0.911} \\
    
\hline
    LibriCount & Accuracy & \textbf{0.46} & 0.422 & 0.451 \\
\hline
    Gunshot Triangulation & Accuracy & 0.911 & 0.917 & \textbf{0.932} \\
\hline
    ESC-50 & Accuracy & 0.334 & 0.325 & \textbf{0.367} \\
\hline
    Mridingham Stroke & Accuracy & 0.967 & \textbf{0.971} & 0.969 \\
\hline
    Mridingham Tonic & Accuracy & \textbf{0.945} & 0.942 & 0.942 \\
\hline    
    \multirow{2}{*}{NSynth Pitch 5h} & Pitch Accuracy & 0.448 & \textbf{0.462} & 0.44 \\
\cline{2-5}   
        & Chroma Accuracy & 0.498 & \textbf{0.516} & 0.48 \\
\hline    
    Speech Commands 5H & Accuracy & 0.974 & \textbf{0.976} & \textbf{0.976} \\
\hline    

\end{tabular}
\end{table*}

There is a difference of at most 3.3\% between the mean and the iterative interpolation algorithm, which is not much, but also not insignificant. It can be seen that generally, the interpolation algorithms out-perform the mean-based approach for generating scene embeddings. And between iterative interpolation and single interpolation, iterative interpolation performs slightly better.


\subsection{Training a Shallower KW-MLP Encoder}
\label{ablation: shallower-kwmlp}

\begin{table*}[ht]
\centering
\caption{Impact of KW-MLP at different encoder depths}
\label{tbl:kwmlp-depth}
\begin{tabular}{ |c|c|c|c|c|  }
\hline
    \multirow{2}{*}{Task} & \multirow{2}{*}{Metric} &   \multicolumn{3}{c|}{KW-MLP Depth} \\
    \cline{3-5}
    & &  8 & 10 & 12 \\
\hline
    Beijing Opera Percussion & Accuracy & \textbf{0.915} & \textbf{0.915} & 0.911 \\
    
\hline
    LibriCount & Accuracy & \textbf{0.475} & 0.432 & 0.451 \\
\hline
    Gunshot Triangulation & Accuracy & 0.860 & 0.917 & \textbf{0.932} \\
\hline
    ESC-50 & Accuracy & \textbf{0.398} & 0.351 & 0.367 \\
\hline
    Mridingham Stroke & Accuracy & 0.970 & \textbf{0.973} & 0.969 \\
\hline
    Mridingham Tonic & Accuracy & \textbf{0.952} & 0.945 & 0.942 \\
\hline    
    \multirow{2}{*}{NSynth Pitch 5h} & Pitch Accuracy & 0.462 & \textbf{0.470} & 0.44 \\
\cline{2-5}   
        & Chroma Accuracy & 0.506 & \textbf{0.516} & 0.48 \\
\hline    
    Speech Commands 5H & Accuracy & 0.969 & \textbf{0.978} & 0.976 \\
\hline    

\end{tabular}
\end{table*}

In the HEAR challenge, we used a KW-MLP encoder with a depth of 12. We wanted to observe whether downstream task performance was proportional to performance on the pretraining task (Speech Commands V2-35) as well as the depth of the encoder.

By varying the number of blocks, $L$,  to 8 and 10, we trained two additional encoder models on Speech Commands V2-35, KW-MLP$_{8}$ and KW-MLP$_{10}$. For reference, KW-MLP$_{12}$ had a top-1 accuracy of 97.56\%, while the KW-MLP$_{8}$ and KW-MLP$_{10}$ we later trained had 97.26\% and 97.35\% respectively.

We then used the evaluation kit provided by HEAR 2021 organizers to evaluate these new encoders. Our comparative results with KW-MLP with varying number of blocks can be observed at Table \ref{tbl:kwmlp-depth}.

With fewer number of blocks, the model is forced to learn more compact representations. Furthermore, the results on other tasks is not proportional to the performance on the pre-training task. This suggests that the high-level features of Speech Commands are detrimental to generalization. In the next section, we attempt to take lower level representations from shallower depths of KWMLP$_{12}$.

\subsection{Generating Embedding from Shallower Depth of KW-MLP}
\label{ablation: shallower-embedding-kwmlp}

\begin{figure}[ht]
\centering
\includegraphics[width=0.6\linewidth]{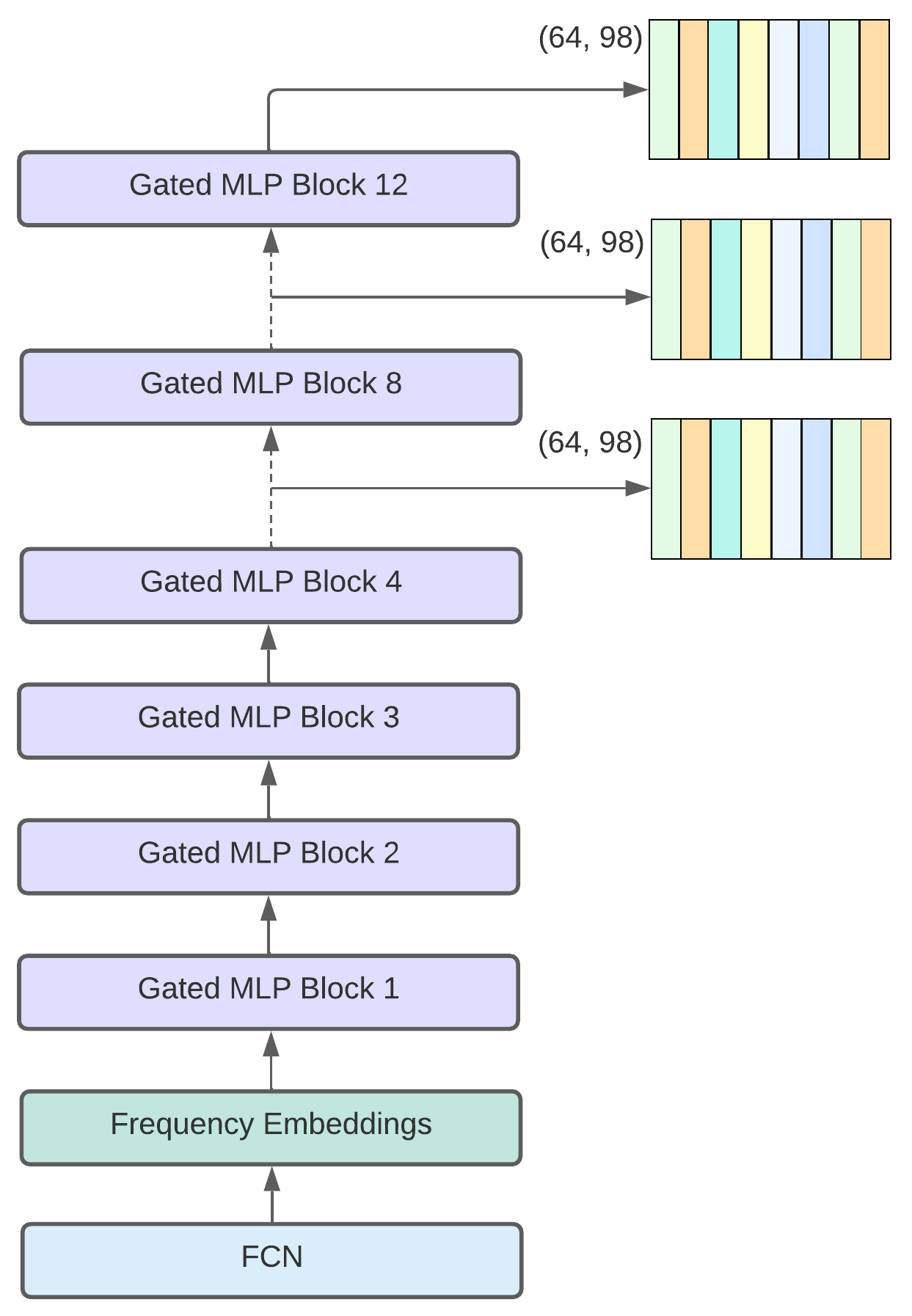}
\caption{We can extract embeddings of the same size from shallower blocks of the model. Note the dotted lines represent subsequent blocks, i.e. between 4 and 8 there is 6 and 7.}
\label{fig:embed_from_shallow}
\end{figure}

\begin{table*}[ht]
\centering
\caption{Generating representations from shallower layers of KW-MLP$_{12}$}
\label{tbl:kwmlp-representation}
\begin{tabular}{ |c|c|c|c|c|  }
\hline
    \multirow{2}{*}{Task} & \multirow{2}{*}{Metric} &   \multicolumn{3}{c|}{KW-MLP$_{12}$  depth} \\
    \cline{3-5}
    & &  4 & 8 & 12 \\
\hline
    Beijing Opera Percussion & Accuracy & \textbf{0.949} & 0.941 & 0.911 \\
    
\hline
    LibriCount & Accuracy & 0.485 & \textbf{0.516} & 0.451 \\
\hline
    Gunshot Triangulation & Accuracy & \textbf{0.976} & 0.943 & 0.932 \\
\hline
    ESC-50 & Accuracy & 0.446 & \textbf{0.485} & 0.367 \\
\hline
    Mridingham Stroke & Accuracy & \textbf{0.977} & \textbf{0.977} & 0.969 \\
\hline
    Mridingham Tonic & Accuracy & \textbf{0.965} & 0.963 & 0.942 \\
\hline    
    \multirow{2}{*}{NSynth Pitch 5h} & Pitch Accuracy & \textbf{0.544} & 0.528 & 0.44 \\
\cline{2-5}   
        & Chroma Accuracy & \textbf{0.576} & 0.562 & 0.48 \\
\hline    
    Speech Commands 5H & Accuracy & 0.786 & 0.932 & \textbf{0.976} \\
\hline    

\end{tabular}
\end{table*}

In deep neural networks, the initial couple of layers capture the low-level features of the data, whereas the final group of layers capture the higher-level features of data. For example, in images, the shallow layers capture the edges, while in the last few layers, the model captures high-level features of the image, such as patterns, objects, and shapes. This phenomenon is also present in our method. We generated representations from encoder depth 4 and 8 from KW-MLP$_{12}$, as shown in Figure \ref{fig:embed_from_shallow}.

From Table \ref{tbl:kwmlp-representation} we observe that in general, representations from shallower depths improve our model's performance, except Speech Commands. In Speech Commands the encoder model at depth 12 outperforms encoder model at depth 4 and 8. This is because the model is trained on a Speech Commands V2-35, and thus losing the learned high-level speech representations hurts performance on Speech Commands V2-12. On the other datasets, the embeddings from shallower layers have significantly better performance, as the representations at those layers are not biased on the Speech Commands dataset. We believe that the intermediate embeddings at encoder depth 4 and 8 better captured the general features of audio data compared to the embeddings at depth 12, which are too strongly geared towards the Speech Commands dataset. 

Comparing with the results at the leaderboard \citep{hearresults}, we obtained better results than the 1st position ranking in Gunshot Triangulation, Mridingham Stroke and Mridingham Tonic tasks. 


\section{Conclusion and Future Work}
\label{sec: conclusion}

Our MLP-based audio representation learning is highly efficient, and performs quite well in certain downstream tasks. However, it also has sub-par performance on other tasks. One of the aims of the HEAR Challenge was to evaluate whether it was possible for models to learn audio representations as holistic as the human ear. From our results, as well as the results of the other submissions, this seems to remain an open question, for now.

In order to enable fast pretraining and have an extremely efficient method, \texttt{kwmlp} has compromised in certain aspects, such as the pretraining dataset or task, and the model size. In future works, we would like to explore self-supervised representation learning methods with similar all-MLP architectures on larger and richer audio corpuses, such as LibriSpeech \citep{librispeech}, AudioSet \citep{audioset} or Synth1B1 \citep{synth1b1}, among others. One fact the HEAR challenge has made apparent is that perhaps it is difficult to learn truly generalized and holistic representations from solely one type of audio data, such as speech.

We hope our work is helpful to future research on audio representation learning, and opens up a new avenue of working with all-MLP models for audio tasks.


\bibliography{pmlr-sample}

\appendix
\pagebreak
\section{Settings and Hyperparameters for KW-MLP}
\label{appendix: kwmlp-hparams}

\begin{table}[ht]
	\centering
	\caption{Overview of Hyper-Parameters and Settings}
	\label{tbl:settings}
	\begin{tabular} {c c c c c}
		\cline{1-2}  \cline{4-5}
		\multicolumn{2}{c}{Training} & & \multicolumn{2}{c}{Augmentation}\\
		\cline{1-2}  \cline{4-5}
		Epochs & 140 &  & \# Time Masks & 2 \\
		Batch Size & 256 & & Time Mask Width & [0, 25] \\
		Optimizer & AdamW & & \# Freq Masks & 2 \\
		Learning Rate & 0.001 & & Freq Mask Width & [0, 7] \\
		Warmup Epochs & 10 \\
		Scheduling & Cosine \\
		
		\cline{1-2}  \cline{4-5}  
		\multicolumn{2}{c}{Regularization} & & \multicolumn{2}{c}{Audio Processing} \\
		\cline{1-2}  \cline{4-5} 
		Label Smoothing & 0.1 & & Sampling Rate & 16000 \\
		Weight Decay & 0.1 &  &  Window Length & 30 ms  \\
		Block Survival Prob. & 0.9 & &  Hop Length & 10 ms \\
		& & &  n\_mfcc & 40  \\
		
		\cline{1-2} 
		\multicolumn{2}{c}{Model} \\
		\cline{1-2} 
		\# Blocks, $L$ & 12/10/8 \\
	    Input Shape & $40\times98$ \\
	    Patch Size & $40\times1$ \\
	    Dim, $d$ & 64 \\
	    Dim Proj. & 256 \\
        \# Classes & 35 
	\end{tabular}
\end{table}

\end{document}